\documentclass[11pt,letterpaper]{article}
\usepackage{arxiv}

\usepackage[utf8]{inputenc}
\usepackage[T1]{fontenc}
\usepackage{hyperref}
\usepackage{url}
\usepackage{booktabs}
\usepackage{amsfonts}
\usepackage{nicefrac}
\usepackage{microtype}
\usepackage{lipsum}
\usepackage{graphicx}

\title{Challenges in Implementing a Recommender System for Historical Research in the Humanities}

\date{\vspace{-5ex}}

\author{
  Florian Atzenhofer-Baumgartner$^{1,2}$ \and
  Bernhard C. Geiger$^{3,4}$ \and
  Christoph Trattner$^5$ \and
  Georg Vogeler$^2$ \and
  Dominik Kowald$^{1,3}$
}

\begin{document}
\maketitle

\noindent
$^1$Institute of Interactive Systems and Data Science, Graz University of Technology, Graz, Austria\\
$^2$Department of Digital Humanities, University of Graz, Graz, Austria\\
$^3$Know Center Research GmbH, Graz, Austria\\
$^4$Signal Processing and Speech Communication Laboratory, Graz University of Technology, Graz, Austria\\
$^5$University of Bergen \& Media Futures, Bergen, Norway

\begingroup
\renewcommand\thefootnote\relax
\footnotetext{Corresponding author: Florian Atzenhofer-Baumgartner, atzenhofer@acm.org}
\endgroup

\vspace{2em}

\begingroup
\renewcommand\thefootnote\relax
\footnotetext{Presented at AltRecSys 2024: The First Workshop on Alternative, Unexpected, and Critical Ideas in Recommendation, October 18, 2024, co-located with the ACM Conference on Recommender Systems 2024 (RecSys 2024), Bari, Italy.}
\endgroup

\begin{abstract}
This extended abstract describes the challenges in implementing recommender systems for digital archives in the humanities, focusing on Monasterium.net, a platform for historical legal documents. We discuss three key aspects: (i) the unique characteristics of so-called charters as items for recommendation, (ii) the complex multi-stakeholder environment, and (iii) the distinct information-seeking behavior of scholars in the humanities. By examining these factors, we aim to contribute to the development of more effective and tailored recommender systems for (digital) humanities research.
\end{abstract}


\vspace{1em}

\section{Introduction}

Recommender systems (RecSys) are ubiquitous in various domains (e.g., entertainment and multimedia~\cite{kowald2022popularity}), enhancing user experience and facilitating content discovery. However, their application in digital humanities venues is underutilized and presents unique challenges. This work focuses on Monasterium.net, a digital archive for charters, to explore three main challenges (see Sections~\ref{s:charter} -- \ref{s:information}) and propose potential directions for future research (see Section~\ref{s:conclusion}).

\section{Charters as Unique Items for Recommender Systems}
\label{s:charter}
Charters, the primary content of Monasterium.net, have distinct characteristics that set them apart from typical items in RecSys. These historical documents record legal actions and provide rich insights into past societies \cite{Slavin_2012}. Unlike commercial products or entertainment content, charters are complex, multifaceted artifacts with varying levels of completeness and authenticity. In addition, they exhibit a specific type of public value as cultural heritage objects.

Charters in Monasterium.net are represented as digitized scans, scholarly editions, and semi-structured data using specialized schemas like the Charter Encoding Initiative (CEI) \cite{Vogeler_2005}. This representation includes diverse metadata such as material properties, production details, authentication methods, and scholarly annotations. The high dimensionality and often sparse population of this metadata pose significant challenges for traditional recommendation algorithms.

\begin{figure}[htb]
    \centering
    \includegraphics[width=0.8\textwidth]{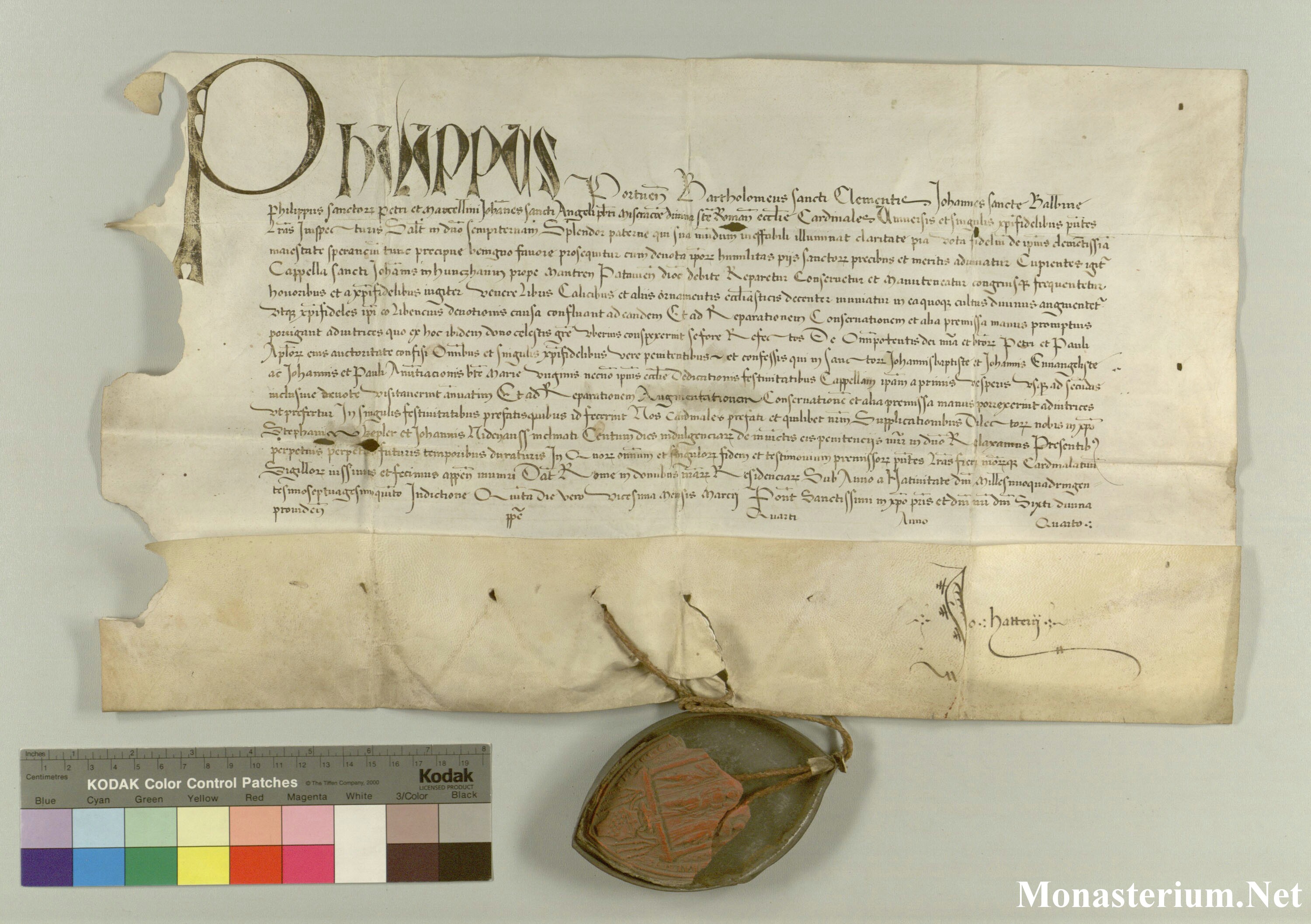}
    \caption{Recto side of a charter in which five Cardinal priests grant a 100-day indulgence for St. John's Chapel in Mautern on the Danube (1475). The document, requested by local petitioners, details specific visitation requirements and contributions; it demonstrates typical medieval ecclesiastical documentation practices and authentication methods. Adapted from \href{https://www.monasterium.net/mom/AT-StiAG/GoettweigOSB/1475\_III\_20/charter}{https://www.monasterium.net/mom/AT-StiAG/GoettweigOSB/1475\_III\_20/charter}.}
    \label{fig:cardinal-charter-1475}
\end{figure}

Furthermore, the historical nature of charters introduces additional complexities. Unlike continuously growing datasets in modern RecSys, the set of historical charters is finite, though new discoveries or digitizations may occasionally expand the corpus. Charters span different historical periods, requiring consideration of temporal relevance in recommendations. Some charters may be identified as forgeries long after their creation, affecting their relevance and interpretation. Many charters contain non-resolvable, missing references of entities, or damaged sections, which requires special handling in data representation and recommendation algorithms. Additionally, charters often contain text in various languages and scripts, which furthers complicates text-based recommendation approaches.

\section{Multi-Stakeholder Environment in Digital Archives}

The implementation of RecSys for Monasterium.net must consider a complex ecosystem of stakeholders in a mostly non-commercial setting, each with distinct values and objectives. Key stakeholders include researchers, primarily historians and other humanities scholars, who seek relevant documents for their studies. They value accuracy, comprehensiveness, and the ability to discover unexpected connections \cite{Nix_Decker_2023}. Content creators, including archivists and editors who contribute to the platform, aim for visibility and recognition of their work \cite{Andro_Saleh_2017}. Platform owners, such as ICARus for Monasterium.net, focus on user satisfaction, engagement, and strategic growth. Funding agencies, who support the digitization and preservation of charters, value the extent of use and impact of the collections.

Balancing these diverse and sometimes conflicting interests presents a significant challenge. For example, recommendations that prioritize popular or well-documented charters might satisfy casual users but could limit the exposure of lesser-known documents that might be crucial for specialized research. Similarly, the needs of content creators for visibility might conflict with researchers' requirements for the most relevant documents, regardless of their popularity or recency of addition to the platform.

\section{Information-Seeking Behavior in Humanities Research}
\label{s:information}
The information-seeking behavior of humanities scholars differs significantly from the user behavior in commercial RecSys. For instance, historians often engage in exploratory searches, seeking to uncover unexpected connections and generate new research questions \cite{Martin_Quan-Haase_2016,Engerer_2021}. Their research process is typically non-linear, involving iterative refinement of queries and exploration of diverse paths, and thus, suggesting incorporating context information such as time or semantic context cues \cite{kowald2015refining,kowald2015forgetting}. 

Moreover, the concept of relevance in the humanities is multifaceted and context-dependent. A charter's relevance may not be immediately apparent and could emerge through the process of investigation. This characteristic challenges traditional relevance metrics used in RecSys. Connected to this, the long-term nature of humanities research also impacts user interaction with the system. Unlike in conventional RecSys, where immediate user feedback is common, the value of a recommendation in humanities research might only become apparent after extended periods of study.

\section{Conclusion and Future Work}
\label{s:conclusion}
Implementing effective RecSys for digital archives like Monasterium.net requires addressing the unique nature of historical documents, navigating a complex multi-stakeholder environment, and accommodating the distinct information-seeking behavior of humanities scholars. Future work should focus on developing specialized algorithms that can handle the complexity and sparsity of charter metadata, designing evaluation frameworks that reflect the diverse stakeholder values, and creating interfaces that support the exploratory and long-term nature of humanities research. These endeavors meet the recent calls of the RecSys community for more detailed user studies and context-aware RecSys \cite{Abdollahpouri_Adomavicius_Burke_Guy_Jannach_Kamishima_Krasnodebski_Pizzato_2020}.
Also, by addressing these challenges, we can enhance the utility of digital archives, enable new discoveries and deeper understanding in historical research. This work not only benefits Monasterium.net, but also provides insights applicable to other digital humanities platforms and cultural heritage institutions \cite{Colavizza_Blanke_Jeurgens_Noordegraaf_2022}.

\vspace{2mm} \noindent \textbf{Acknowledgements}. 
The work presented in this paper has been supported by the ERC Advanced Grant project (101019327) ``From Digital to Distant Diplomatics''. Additionally, this research was supported by the Know Center Research GmbH within the COMET — Competence Centers for Excellent Technologies Programme, funded by bmvit, bmdw, FFG, and SFG.

\bibliographystyle{unsrt}
\bibliography{bibliography}

\end{document}